\documentclass[aip, preprint, showpacs, superscriptaddress, 12pt]{revtex4-1}

\usepackage{amsmath,amssymb,graphicx,natbib}
\usepackage{graphicx}
\usepackage{dcolumn}
\usepackage{bm}
\usepackage{epstopdf}

\begin{document}

\title{Protection layers on a superconducting microwave resonator toward a hybrid quantum system}

\author{Jongmin Lee}
\email{jongmin.lee@sandia.gov}
\affiliation{Joint Quantum Institute, Department of Physics, University of
Maryland and National Institute of Standards and Technology, College Park, Maryland 20742, USA}
\affiliation{Sandia National Laboratories, Albuquerque, New Mexico 87123, USA}

\author{Dong Hun Park}
\email{leomac@umd.edu}
\affiliation{Department of Electrical and Computer Engineering, University of Maryland, College Park, Maryland 20742, USA}

\begin{abstract}
We propose a protection scheme of a superconducting microwave resonator to realize a hybrid quantum system, where cold neutral atoms are coupled with a single microwave photon through magnetic dipole interaction at an interface inductor. The evanescent field atom trap such as a waveguide/nanofiber atom trap, brings both surface-scattered photons and absorption-induced broadband blackbody radiation which result in quasiparticles and a low quality factor at the resonator. A proposed multiband protection layer consists of pairs of two dielectric layers and a thin nanogrid conductive dielectric layer above the interface inductor. We show numerical simulations of quality factors and reflection/absorption spectra, indicating that the proposed multilayer structure can protect a lumped-element microwave resonator from optical photons and blackbody radiation while maintaining a reasonably high quality factor.
\end{abstract}

\maketitle


\section{Introduction}
Hybrid quantum systems~\cite{Verdu09, Hafezi12, Pritchard13, Regal14, Jessen13, Minniberger13} have been studied for coherent quantum interfaces and quantum information. A particular implementation~\cite{Verdu09} of a superconducting microwave resonator (SMR) and cold neutral atoms with magnetic dipole coupling is promising due to the scalability and the fast gate operation time of superconducting microwave (SM) circuits and the long coherence time of trapped atoms. This hybrid quantum system requires magnetic/optical trappings of cold neutral atoms at the proximity of a SMR, but induced vortices by alternating current (AC) magnetic fields or optically excited quasiparticles by scattered photons recede the performance of a SMR.

\begin{figure}
\centering\includegraphics[width=0.8\textwidth]{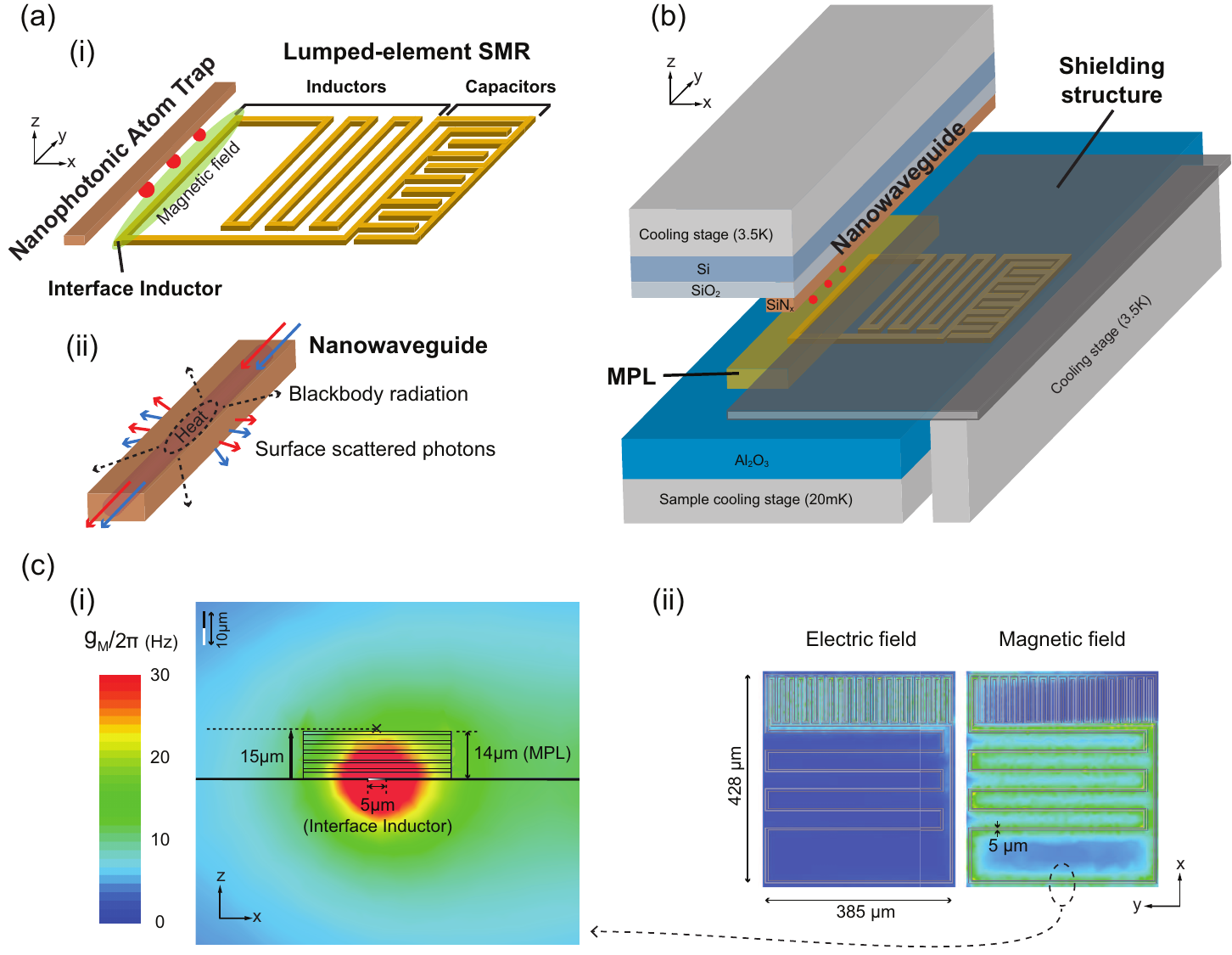}
\caption{The concept of a multiband protection layer (MPL) for a hybrid quantum system. (a) Hybrid quantum system with a lumped-element SMR and a nanophotonic atom trap. (i) Evanescent field trapped atoms coupled to the SMR through magnetic dipole interaction (3D-plot, not-to-scale). (ii) Surface-scattered photons and blackbody radiation from a nanophotonic atom trap create quasiparticles and reduce a Q-factor. (3D-plot, not-to-scale).  The single-photon regime requires to maintain a lower thermal occupation ($T \simeq 20\,\mathrm{mK}$). (b) The protection scheme (3D-plot, not-to-scale). The MPL with a floated shielding structure~\cite{Barends11} protect both inductors and capacitors. (c) The magnetic coupling strength (3D finite element method simulation). (i) The single-photon single-atom Rabi frequency $g_{M}$ according to the proximity (cross-sectional view, xz plane); $g_{M}$ is plotted in the range of $2 \pi \cdot 0\,\textendash\,30\,\mathrm{Hz}$, where $g_{M}|_{z = 10\,\mu\mathrm{m}} \simeq 2\pi \cdot 30\,\mathrm{Hz}$. (ii) Electric fields and magnetic fields of a lumped-element SMR (top view, xy plane).}
\label{fig_concept}
\end{figure}

A proposed hybrid quantum system~\cite{Hafezi12} exploits a lumped-element SMR~\cite{ZKim11} and a nanophotonic atom trap~\cite{Vetsch10, Lee15, Lee13, Yang15} (Fig.~\ref{fig_concept}). A lumped-element SMR has a single resonant frequency and a compact size as submillimeter square, and inductance and capacitance determine a SMR resonant frequency. Inductors/capacitors with oscillating current flows (charge distributions) store energy in magnetic/electric fields, and the magnetic/electric energy is equally separated at inductors/capacitors. Unlike a superconducting coplanar waveguide resonator~\cite{Verdu09}, radio frequency (RF) magnetic/electric fields of a lumped-element SMR in the near-field are localized at inductors/capacitors. Magnetic dipole interaction of a SMR photon and trapped atoms occurs through the interface inductor.

\begin{figure}
\centering\includegraphics[width=0.8\textwidth]{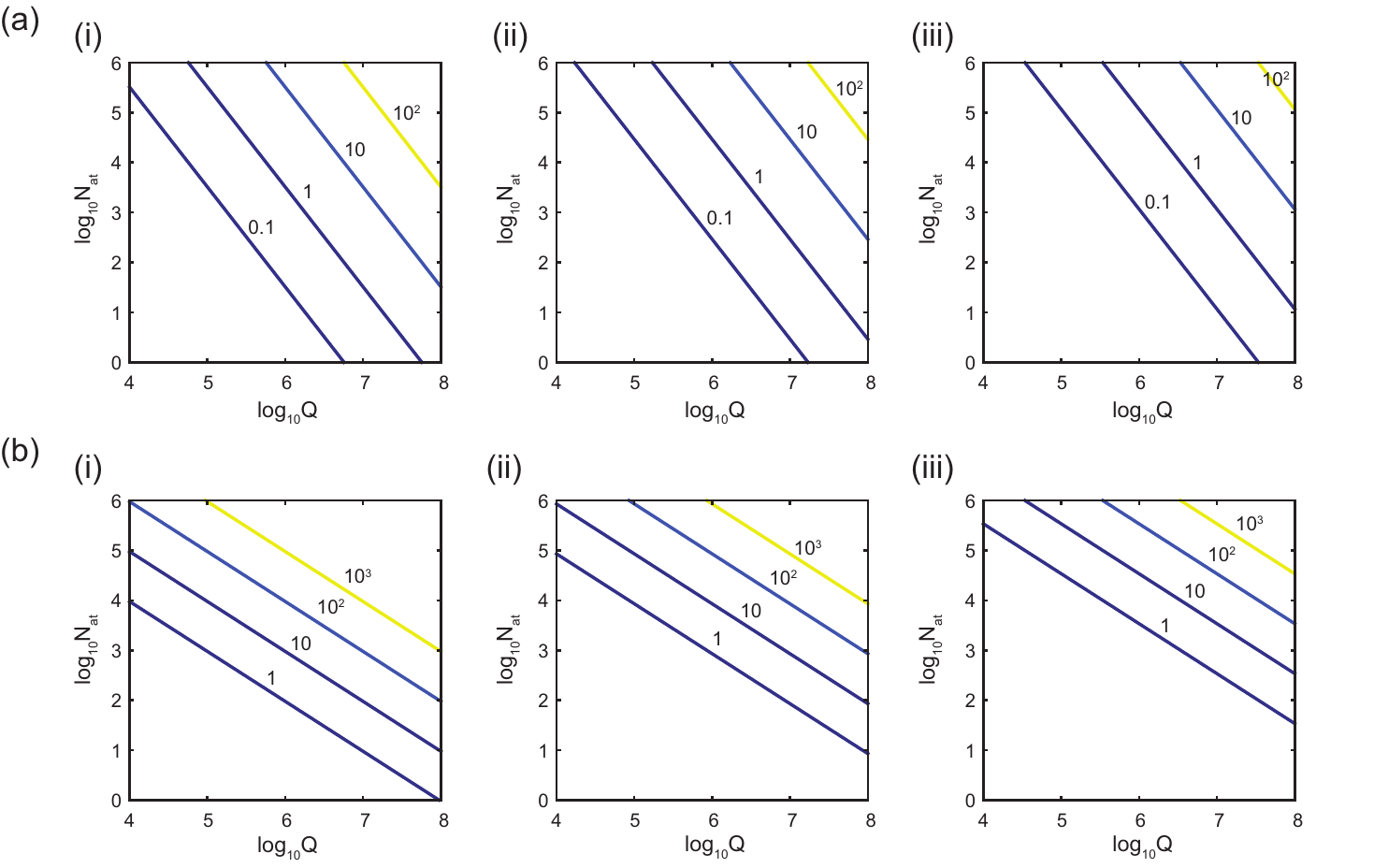}
\caption{The validation of a hybrid quantum system in the single-photon regime. (a) $\sqrt{N_{at}} \cdot g_{M}/(\kappa/2) > 1$ for the multiple coherent exchanges between a single SMR photon and many atoms within a cavity photon lifetime. (i) $z = 5\,\mu\mathrm{m}$ (ii) $z = 15\,\mu\mathrm{m}$ (iii) $z = 30\,\mu\mathrm{m}$, where $\kappa =\,\omega_{MW}/\mathrm{Q}$; $\omega_{MW} = 2\pi \cdot 6.835\,\mathrm{GHz}$. (b) $N_{at} \cdot g_{M}^2/(\kappa \cdot \gamma) \gg 1$ for the coherent collective coupling within sufficient lifetimes of atoms and photons. (i) $z = 5\,\mu\mathrm{m}$ (ii) $z = 15\,\mu\mathrm{m}$ (iii) $z = 30\,\mu\mathrm{m}$, where $\gamma = 2\pi \cdot 50\,\mathrm{Hz}$ mostly determined by the trap lifetime; $g_{M}|_{z} \simeq 2 \pi \cdot (60,\, 20,\, 10)\,\mathrm{Hz}$ for $z$ = (5,\, 15,\, 30)\,$\mu\mathrm{m}$, respectively. In case of $N_{at} = 10^4$ and $\mathrm{Q} = 10^6$, $\sqrt{N_{at}} \cdot g_{M}|_{z=15\,\mu\mathrm{m}} = 2 \pi \cdot 2\,\mathrm{kHz} \lesssim 2\pi \cdot 3.4\,\mathrm{kHz} = \kappa/2$ and $N_{at} \cdot g_{M}^2|_{z=15\,\mu\mathrm{m}}/(\kappa \cdot \gamma) \simeq 12 > 1$. In case of $N_{at} = 10^6$ and $\mathrm{Q} = 10^6$, $\sqrt{N_{at}} \cdot g_{M}|_{z=15\,\mu\mathrm{m}} = 2\pi \cdot 20\,\mathrm{kHz} > 2\pi \cdot 3.4\,\mathrm{kHz} = \kappa/2$ and $N_{at} \cdot g_{M}^2|_{z=15\,\mu\mathrm{m}}/(\kappa \cdot \gamma) \simeq 1.2 \times 10^3 \gg 1$.}
\label{fig_HQS}
\end{figure}

A hybrid quantum system necessitates commensurable single-photon (single-atom) magnetic coupling and single-photon collective (many-atom) magnetic coupling. The magnetic field of a single SMR photon can be estimated as $\langle \vec{B_{1}} \rangle = \langle \vec{B} \rangle / \sqrt{\langle n_{ph} \rangle}$. The single-photon single-atom Rabi frequency is 
\begin{equation}
g_{M} = \frac{\vec{\mu}_{B} \cdot \langle \vec{B_{1}} \rangle} {\hbar},
\end{equation}
where $\vec{\mu}_{B}$ is the magnetic dipole moment of neutral atoms ($^{87}$Rb) in hyperfine ground states. For the coherent interface between a single SMR photon and many atoms, a hybrid quantum system requires the multiple coherent exchanges between a single SMR photon and many atoms within a cavity photon lifetime such as $\sqrt{N_{at}} \cdot g_{M} > \kappa/2$ \,(Fig.~\ref{fig_HQS}(a)), where $N_{at}$ is an atom number; $\kappa$ is a cavity decay rate. The coherent collective coupling~\cite{Thompson92, Tuchman06, Lee14} also entails sufficient lifetimes of atoms and photons such as the magnetic cooperativity $N_{at} \cdot g_{M}^2/(\kappa \cdot \gamma) \gg 1$\, (Fig.~\ref{fig_HQS}(b)), where $\omega_{MW}$ is a SMR resonant frequency  and $\gamma$ is an atomic decay rate of a hyperfine ground state atom in the evanescent field atom trap. The Hamiltonian of collective atom-photon interaction\cite{Verdu09} is represented as
\begin{equation}
\hat{H} = \hbar \omega_{MW} \hat{a}^{\dagger}\hat{a} + \hbar \omega_{a} \tilde{\pi}^{\dagger} \tilde{\pi} + \hbar g_{eff} (\tilde{\pi}^{\dagger}\hat{a} + \hat{a}^{\dagger}\tilde{\pi}),
\end{equation}
where $g_{eff} = \sqrt{N_{at}} \cdot g_{M}$ is the single-photon collective magnetic coupling; $\hat{a}^{\dagger}$ ($\hat{a}$), the creation (annihilation) operator of SM photons; $\omega_{a}$, an atomic transition frequency of $^{87}$Rb (D2) hyperfine ground states; $\tilde{\pi} = 1/\sqrt{N_{at}}\cdot \sum_{i=1}^{N_{at}} \hat{\pi_{i}}$, the collective atomic operator; $\hat{\pi}_{i}^{\dagger}$, the excitation operator of $i$-th atom.

A nanophotonic atom trap uses optical potentials to confine atoms along the transverse directions with the differential decay lengths of two-color traveling evanescent wave fields as well as atoms along the propagation direction with a standing wave of red-detuned fields~\cite{Vetsch10, Lee15, Lee13, Yang15}. For a single atomic transition ($\omega_{0}$), the induced light shift (AC Stark shift) creates an optical potential. This potential~\cite{Grimm00} for a large detuning ($\Delta_{trap}=\omega-\omega_{0}$) is 
\begin{equation}
U_{opt}(\mathbf{r})= \frac{3\pi c^2}{2\omega_{0}^{3}} \frac{\Gamma}{\Delta_{trap}} I(\mathbf{r}),
\end{equation}
where $I(\mathbf{r})$ is the laser intensity and $\Gamma$ is the spontaneous decay rate of the excited state. For $^{87}\textrm{Rb}$ atoms ($\lambda_{0} = 780\,\textrm{nm}$), a 750\,nm blue-detuned trapping beam creates a repulsive potential ($\Delta_{blue}>0$), and a 1064\,nm red-detuned trapping beam produces an attractive potential ($\Delta_{red}<0$). For trapping atoms near the surface ($\sim$200\,nm) of a dielectric, the blue-detuned beam needs to compensate the attractive van der Waals potential. Approximating the surface to be an infinite dielectric, the van der Waals potential is $U_{vdW}(y) = - C_{vdW} \cdot y^{-3}$, where $y$ is the distance from the waveguide surface, and $C_{vdW}$ is determined from atomic dipole transition. The total potential along the vertical direction of the waveguide is
\begin{equation}
U_{tot} = U_{blue} + U_{red} + U_{vdW}.
\end{equation}
The evanescent field atom trap can confine and transfer atoms close to a SMR without AC magnetic fields, but surface-scattered photons and absorption-induced blackbody radiation (BBR) higher than a superconducting bandgap can dissipate superconductivity~\cite{Barends11}.

Maintaining a high Q-factor is crucial for the multiple coherent exchanges and the collective magnetic coupling between evanescent field trapped atoms and a cavity mode. A high Q-factor is essential to implement the coherent interface in a hybrid quantum system. Satisfying those conditions, the multiband protection layer (MPL) is designed to maintain a Q-factor from optically and thermally excited quasiparticles created by two-color near-infrared (NIR) optical trapping photons and a broadband mid-infrared (MIR) BBR (Appendix \ref{Appx_SC_sensitivity}). 

The MPL brings additional loss mechanisms such as dielectric loss tangent and two-level system (TLS) dissipation. In a lumped-element SMR, dielectric loss tangent and TLS dissipation at the interface inductor affects a Q-factor much less than those at capacitors; inductors with localized magnetic fields are less sensitive to dielectrics and electric dipole moments compared to capacitors with localized electric fields, and the MPL is located above the interface inductor where the localized magnetic field couples to trapped atoms. Capacitors and other inductors located away from the interface are protected against scattered photons and BBR by a floated shielding structure.

In this paper, we study a protection scheme to address the problems that arise from coupling a lumped-element SMR with evanescent field trapped atoms. We design the MPLs that suppress the transmissions of two NIR optical trapping fields and a broadband MIR BBR to protect the resonator and maintain a reasonable Q-factor toward the realization of a hybrid quantum system with neutral atoms.

\section{Design and Simulation}

In this section, we discuss an MPL located on the interface inductor of a lumped-element SMR for the realization of a hybrid quantum system with the evanescent field atom trap. It consists of dielectric pairs with a thin conductive dielectric layer to reduce optical and thermal quasiparticle excitations and to maintain a single-photon magnetic coupling strength. The MPL should be designed to suppress the multiband transmissions of two-color surface-scattered NIR photons and an absorption-induced broad MIR blackbody radiation. The angle dependency of the scattered photons is studied because the transmission of oblique incident fields through the MPL may affect the hybrid quantum system. The maximum allowable BBR temperature on the interface inductor area without the MPL is estimated based on the allowable power to the resonator in the steady state. The surface scattering loss from the structure and the absorption-induced heat flux from the extinction coefficient are also calculated in case of the nanophotonic atom trap~\cite{Vetsch10, Lee15, Lee13, Yang15}. We discuss the attenuation of the magnetic field with varying MPL thickness because the magnetic dipole coupling is determined by the back-transmitted magnetic field through the MPL. The reduced Q-factor caused by dielectric loss and two-level system dissipation of the optimal MPL is estimated by a realistic range of a loss tangent using finite element method (FEM) simulation.

\begin{figure}
\centering\includegraphics[width=0.8\textwidth]{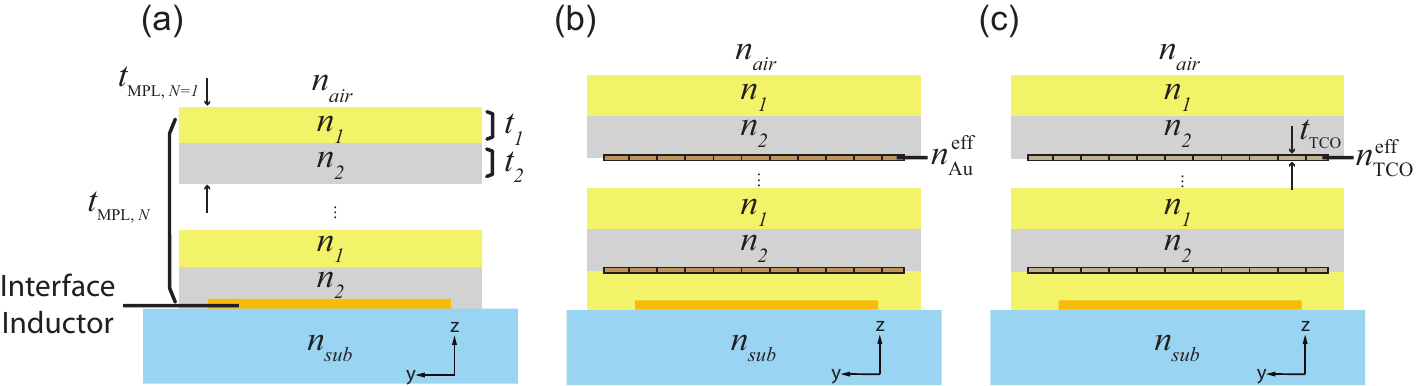}
\caption{Cross-sectional view of the MPL on the interface inductor (yz plane, not-to-scale). (a) The MPL layer with dielectric pairs. (b) The MPL with a thin nanogrid metallic layer. (c) The MPL with a thin nanogrid conductive dielectric layer, where $n_{1}$, $n_{2}$, $n_{\mathrm{Au}}^{\mathrm{eff}}$, and $n_{\mathrm{TCO}}^{\mathrm{eff}}$ are refractive indices of dielectric pairs, a thin metallic layer (Au), and a thin conductive dielectric layer ($\mathrm{TiO_2}$); the number of pairs $N$; dielectric layer thickness $t_{1}$ and $t_{2}$; single MPL thickness $t_{_{\mathrm{MPL},\,N=1}}$; total MPL thickness $t_{_{\mathrm{MPL},\,N}} = N \cdot\,t_{_{\mathrm{MPL},\,N=1}} = N \cdot\,\left(t_{1}+t_{2}+t_{TCO}\right)$ excluding substrate; transparent conductive oxide (TCO) with its thickness $t_{TCO}$. The thin nanogrid conductive dielectric layer enhances BBR suppression, and sub-micrometer grid-structures with a lower bulk conductivity reduce conductive and induced-current losses.}
\label{fig_several_Bragg}
\end{figure}

The MPL is required to achieve high reflectance and low transmission for the multiband spectrum, and the protection of BBR is important because the optical absorption at the waveguide/nanofiber causes heat and generating BBR. We consider three different types of MPLs, (i) the MPL of dielectric pairs, (ii) the MPL with a thin nanogrid metallic layer (Au), and (iii) the MPL with a thin nanogrid conductive dielectric layer as illustrated in Fig.~\ref{fig_several_Bragg}. An exemplary nanowaveguide geometry based on Refs~\cite{Lee13, Yang15} is used.

Firstly, we design a reflector with dielectric pairs (R$>$99.9\,\%) for the spectrum of two-color evanescent atom trap beams (750\,nm and 1064\,nm), considering the Bragg condition $\lambda/(4 n_{i})$, where $n_{i}$ is an index of refraction. The bandwidth of stopbands is linearly proportional to the index difference $n_{1} - n_{2}$ between two neighboring materials~\cite{Pochi}. Secondly, the MIR reflector with high reflection and wide bandwidths is obtained by high index difference~\cite{Heiss01}; the high index difference is practically limited by the optical properties of materials. A thin MPL thickness is required for a better single-photon magnetic coupling strength as shown in Fig.~\ref{fig_concept}~(c-i) and Fig.~\ref{fig_HQS} because the magnetic field proportional to the square root of a cavity photon number decays with distance. The design requires a broadband MIR reflector with a good single-photon magnetic coupling $g_{M}$.

Assume a multilayer structure as illustrated in Fig.~\ref{fig_several_Bragg} (a). Based on Airy's formula, the reflection coefficient for the multilayer structure as illustrated in Fig.~\ref{fig_several_Bragg} can be calculated~\cite{Pochi}. When each layer thickness ($t_{i}$) is designed to be $\lambda/(4n_{i}),$ reflectivity from a conventional design of reflector can be expressed as
\begin{equation}
R=\left[\frac{1-\frac{n_{sub}}{n_{air}}\left(\frac{n_{1}}{n_{2}}\right)^{2N}}{1+\frac{n_{sub}}{n_{air}}\left(\frac{n_{1}}{n_{2}}\right)^{2N}}\right]^{2},
\label{eq_R}
\end{equation}
where $n_{1}$ and $n_{2}$ ($n_{1}>n_{2})$ are refractive indices of the repeating layers and $n_{air}$, $n_{sub}$, $N$ are refractive indices of air and substrate, and the number of pairs, respectively. Total MPL thickness $t_{_{\mathrm{MPL},\,N}}$ is determined by the number of pairs $N$. One may decide $N$ depending on design limitations such as material indices and reflectance allowed for desired performance. From Eq.~(\ref{eq_R}), we have
\begin{equation}
N =\frac{1}{2}\frac{log\left(\frac{1-\sqrt{R}}{1+\sqrt{R}}\right)+log\left(\frac{n_{air}}{n_{sub}}\right)}{log\left(\frac{n_{1}}{n_{2}}\right)}.
\label{eq_N}
\end{equation}
From Eqs. (\ref{eq_R}) and (\ref{eq_N}), we note that the reflectivity $R$ approaches unity for a large $N$. Special care must be taken to note that the substrate index affects reflectivity.

\begin{figure}
\centering\includegraphics[width=0.8\textwidth]{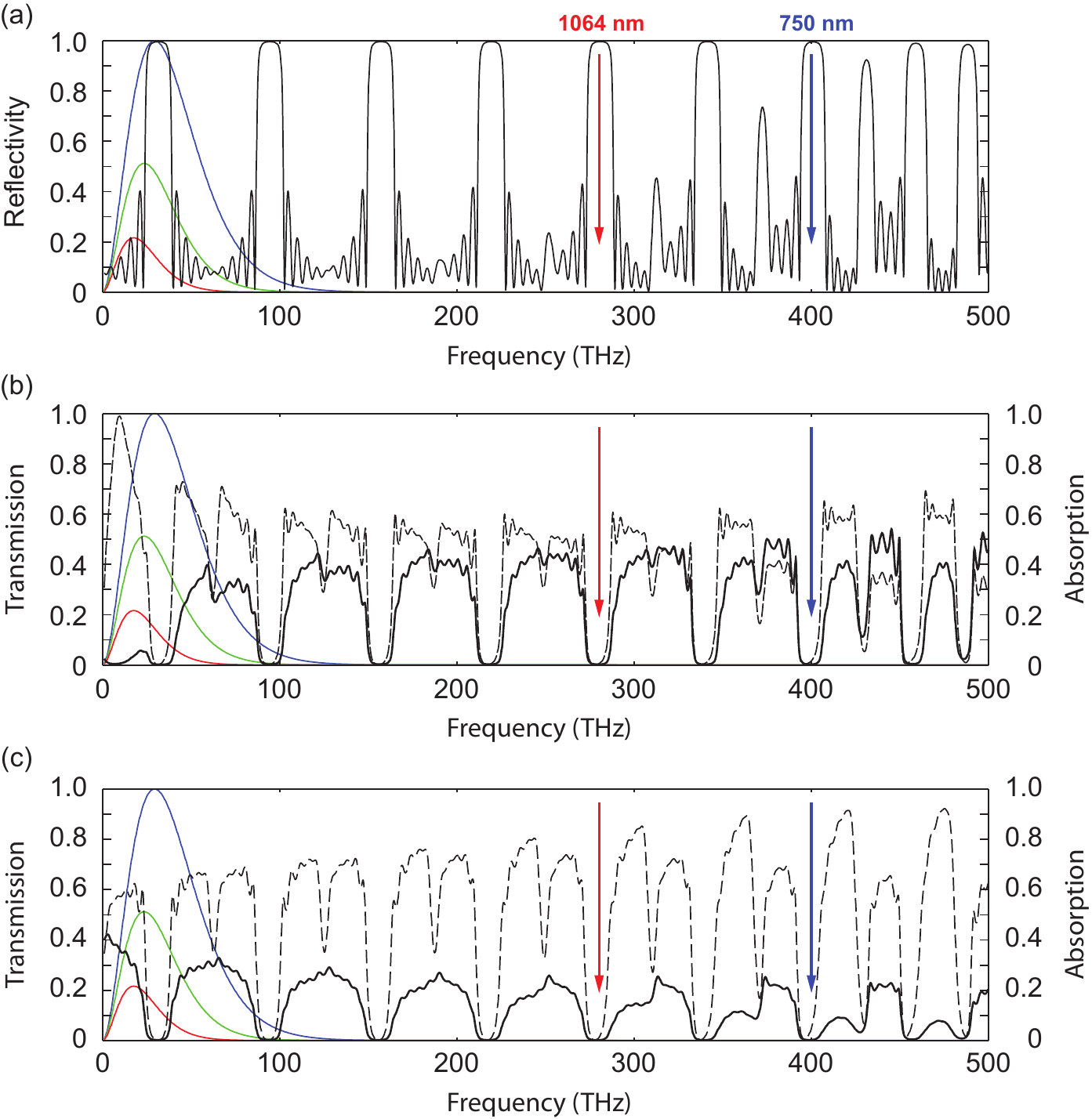}
\caption{The reflectivity and transmission of the MPLs ($t_{_{\mathrm{MPL},\,N = 5}} \simeq 14\,\mu\mathrm{m}$). (a) The MPL of dielectric pairs with $t_{i} = 7.1\lambda/(4n_{i})$ instead of $\lambda/(4n_{i})$; the MPL reflectivity (black) shows the partial prevention of a BBR spectrum. The MPL transmission is $1 - \mathrm{reflectivity}$ due to no absorption at dielectric pairs. (b) The MPL with a thin nanogrid metallic layer; the MPL transmission (black) and absorption (black dash) shows the prevention of a broad BBR spectrum. (c) The MPL with a thin nanogrid conductive dielectric layer; the MPL transmission (black) and absorption (black dash) shows the reduction/prevention of a broad BBR spectrum, and this configuration is an optimal protective solution against two-color NIR photons and a broadband MIR BBR to maintain a Q-factor (Fig.~\ref{fig_Q_factor}) for the single-photon collective coupling. The MPL absorption (Fig.~\ref{fig_Bragg_reflectors} (b-c)) can be harmful because it can increase thermal background. The absorption of the MPL with a thin nanogrid conductive dielectric layer is studied as shown in Fig.~\ref{fig_abs_Bragg}. The BBR spectrums of 300 (red), 400 (green), and 500\,K (blue) are considered for a nanofiber atom trap with a significant absorption-induced BBR compared to a waveguide atom trap. Two-color evanescent field atom trap beams are 1064\,nm (red arrow) and 750\,nm (blue arrow).}
\label{fig_Bragg_reflectors}
\end{figure}

The high index material for the first layer can be either silicon nitride ($\mathrm{SiN_x}$, $n_{\mathrm{SiN_x}} = 1.9\,\textendash\,2.4$) or titanium dioxide ($\mathrm{TiO_2}$, $n_{\mathrm{TiO_{2}}}=2.5\,\textendash\,2.9$)~\cite{Shuaib11,Persano06}. The second layer can be either magnesium fluoride ($\mathrm{MgF_{2}}$,  $n_{\mathrm{MgF_{2}}}=1.38$) or a low index polymer material such as Cytop ($\mathrm{n_{cytop}=1.3}$)~\cite{cytop}. By making the operating resonant wavelength odd number multiples, $\lambda=m\times\lambda_{res}$ ($\lambda_{res}$: resonant wavelength, $m$: odd integer), we achieve smaller free spectral range (FSR) and have $\sim99$\,\% reflection at both wavelengths of interest. The thickness of each layer is $t_{i} = m\lambda / (4n_{i})$. The odd number $m$ can be slightly tweaked to cover two-color trapping beams. In this example, when $m=7.1$, the reflectances reach $\sim\,1$ at both wavelengths. Figure~\ref{fig_Bragg_reflectors} (a) shows calculated reflectance as a function of frequency, using $\mathrm{SiN_x}$ and Cytop, and $N = 5$. Unlike the highly reflective properties at two NIR wavelengths, the reflectance at MIR range $5\,\textendash\,100\,\mu\textrm{m}$ does not cover broadband emission because of the nature of a broad BBR spectrum. As shown in Fig.~\ref{fig_Bragg_reflectors} (a), the BBR spectrums corresponding to assumed blackbody temperatures 300, 400, and 500\,K range from 0 to 200\,THz. To see how much energy can be reflected or transmitted, we define reflected and transmitted BBR efficiencies using the overlap integral in the form
\begin{equation}
\mathrm{R_{BBR}}=\frac{\int_{0}^{\infty} S_{BBR}(\nu,T) \cdot |r(\nu)|^{2}\,d\nu}{ \int_{0}^{\infty} S_{BBR}(\nu,T) \,d\nu},\,
\mathrm{T_{BBR}}=\frac{\int_{0}^{\infty} S_{BBR}(\nu,T) \cdot |t(\nu)|^{2}\,d\nu}{ \int_{0}^{\infty} S_{BBR}(\nu,T) \,d\nu},
\label{eq_Overlap}
\end{equation}
where $S_{BBR}$ is a BBR spectrum as a function of frequency and temperature, and $\mathrm{r(\nu)}$ and $\mathrm{t(\nu)}$ are reflection and transmission coefficients, respectively.

The small mode area of a nanowaveguide induces an intensified optical field, and it generates heat in the nanowaveguide core ($\mathrm{SiN_x}$), $\mathrm{SiO_2}$ layer, and its Si substrate. In case of an optical nanofiber~\cite{Vetsch10, Lee15}, the BBR from a sub-wavelength structure has been studied using computational electrodynamics~\cite{Kruger11}. Based on Stefan-Boltzmann law, the power radiated from BBR is  $P = \varepsilon(\omega, T)\cdot\sigma \cdot A \cdot T^4$, where $\varepsilon(\omega, T)$ is an emissivity as a function of a radiation frequency and a temperature. $A$, $\sigma$, and $T$ are the surface area of an object, Stefan-Boltzmann constant, and a temperature of an object, respectively. Without the MPL, the maximum allowable BBR temperature at the interface of a lumped-element SMR is estimated to be $T_{0} \simeq 27.2\,\mathrm{K}$, taking into account the area of the interface inductor and the allowable power to the SMR in the steady-state. The MPL could reduce BBR transmitted to the interface inductor and increase this $T_{0}$, but the temperature dependence of Stefan-Boltzmann law limits the BBR protection on the interface inductor. Therefore, it is difficult to supress the BBR radiated power ($T > 100\,\mathrm{K}$) in the single-photon operation that requires a thin MPL for a good single-photon magnetic coupling.

The MPL of five dielectric pairs ($N = 5$, $t_{_{\mathrm{MPL}},\,N=5}$ = 14\,$\mu$m) show high reflection, almost 100\,\% ($\mathrm{transmission} \simeq 10^{-3}$) at both NIR photons as shown in Fig.~\ref{fig_Bragg_reflectors} (a), whereas the transmitted BBR efficiency $\mathrm{T_{BBR}}$ defined in Eq.~(\ref{eq_Overlap}) is estimated to be 27, 30, and 39\,\% at 300, 400, and 500\,K, respectively. Addition of a thin, nanogrid metallic or conductive dielectric layer in each pair of layers improves the efficiency of protection, suppressing the MPL transmission of a broad BBR spectrum~\cite{Corrigan12} as shown in Fig.~\ref{fig_several_Bragg} (b-c). Thin metal such as Au can be easily deposited and patterned on any polymer and semiconductor surfaces.

We show numerical simulation in case of adding repetitive 2\,nm nanogrid Au layer below $n_2$ as shown in Fig.~\ref{fig_Bragg_reflectors} (b). Note that the reflectivities are still close to 1 at both wavelengths, and the transmitted BBR efficiency $\mathrm{T_{BBR}}$ calculated from Eq.~(\ref{eq_Overlap}) is less than 3\,\% until 500\,K. Here, nanogrid structures are used because planar metal structure can cause magnetic-field-induced loss resulting from eddy currents. Patterned nanogrids smaller than operating wavelength can drastically minimize this loss because of localized magnetic vortices. Optical properties of a subwavelength nanogrid structure can be approximated by Maxwell-Garnett's effective medium theory~\cite{Maxwell-Garnett04} in the form
\begin{equation}
n_{eff}\,=\,n_m\sqrt{\,\frac{2(1 - \delta)n_m^2 + (1 + 2\delta)n^2}{(2 + \delta)n_m^2 + (1 - \delta)n^2}},
\label{Maxwell-Garnett}
\end{equation}
where $n_m$, $n$, and $\delta$ are the refractive indices of the metal, the dielectric, and the volume portion of dielectric material, respectively.

\begin{figure}
\centering\includegraphics[width=0.8\textwidth]{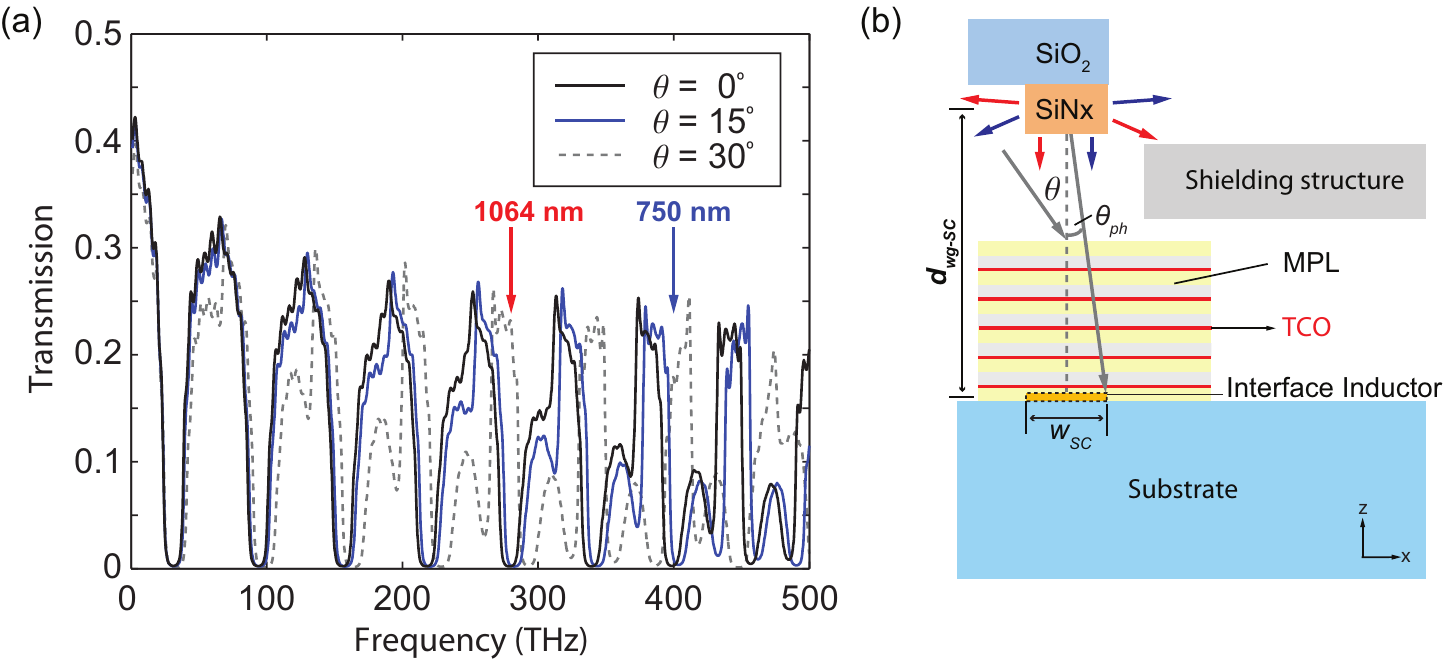}
\caption{The transmission of oblique incident fields through the MPL ($t_{_{\mathrm{MPL},\,N = 5}} \simeq 14\,\mu\mathrm{m}$). (a) The transmission of oblique incident fields for the MPL with a thin nanogrid conductive dielectric layer ($\mathrm{TiO_2}$). The incident angles of $0^{\circ}$ (black), $15^{\circ}$ (blue), $30^{\circ}$ (grey dot) are represented; Fig.~\ref{fig_Bragg_reflectors} (c) is the case of an $0^{\circ}$ incident angle; two-color evanescent field atom trap beams are 1064\,nm (red arrow) and 750\,nm (blue arrow). (b) Cross-sectional view of design geometry (xz plane, not-to-scale). The geometric scattering factors are defined as $\eta_{wg} = 2 \theta_{ph}/180^{\circ}$ (waveguide) and $\eta_{nf} = 2 \theta_{ph}/360^{\circ}$ (nanofiber), assuming the bottom-surface scattering of the waveguide reflected back to the top-surface, where a scattering angle is $\theta_{ph} = \mathrm{tan^{-1}}(0.5 \, w_{sc}/d_{wg\textnormal{-}sc})$; the width of a SMR circuit line, $w_{sc} = 5\,\mu\mathrm{m}$; the distance from a waveguide/nanofiber to the resonator, $d_{wg\textnormal{-}sc}$. For example, geometric scattering factors are $\eta_{wg}|_{15\,\mu\mathrm{m}} \simeq 0.1$ and $\eta_{nf}|_{15\,\mu\mathrm{m}} \simeq 0.05$ with $d_{wg\textnormal{-}sc} = 15\,\mu\mathrm{m}$.}
\label{fig_oblique}
\end{figure}

Figure~\ref{fig_oblique} shows the angle dependency of the MPL from DC to 500\,THz. The light scattering and the BBR can be incident obliquely at a certain angle of incidence. The MIR region does not change much, but optical region shows small variation depending on the incident angle. Most waves from the nanowaveguide are emitted vertically to the MPL and reflected/absorbed, but the oblique wave will partly pass through the MPL. Note that transmissions are still acceptable within $\pm15 ^{\circ}$ ranges.

Both surface scattering loss and absorption-induced heat of a waveguide/nanofiber are the main sources of quasiparticle creation. We estimate the surface scattering loss per surface roughness variance ($\sigma^2$) based on waveguide geometry (Appendix \ref{Appx_Est}). The scattering loss of a strongly guided mode at 1064\,nm light is $\sim$\,0.05\,(dB/cm)/$\mathrm{nm^2}$ ~\cite{Payne94, Barwicz05, Ciminelli09}. The top-surface scattering power per unit length is $\sim$\,340\,$\mu$W/cm for 30\,mW input (assuming $\sigma^2_{top}$ = 1\,$\mathrm{nm}^2$). The side-surface scattering ($\sigma^2_{side} \gg \sigma^2_{top}$) is much higher because of dry etching process, which is blocked by the floated shielding structure (see Fig.~\ref{fig_concept} (b)). This scattering line source with a scattering angle directly degrades the SMR as shown in Fig.~\ref{fig_oblique}. 

The geometric scattering factors of a waveguide (nanofiber) such as $\eta_{wg}$ ($\eta_{nf}$) is defined defined in the caption of Fig.~\ref{fig_oblique}. Therefore, the scattering line source on the MPL above the interface inductor (5\,$\mu$m\,$\times$\,385\,$\mu$m) is reduced to 1.34\,nW by $\eta_{wg}|_{15\,\mu\mathrm{m}}$ and the MPL of $10^{-3}$ transmission ($t_{_{\mathrm{MPL}},\,N=5} = 5 \cdot t_{_{\mathrm{MPL},\,N=1}} \simeq 14\,\mu\mathrm{m}$) with the validation of a hybrid quantum system at $15\,\mathrm{\mu m}$ (see Fig.~\ref{fig_HQS} (a-ii) and (b-ii)). For an optical nanofiber with $2.6\,\times\,10^{-4}$dB/cm propagation loss~\cite{Hoffman14}, the scattering line source (1.8\,$\mu$W/cm out of 30\,mW input) on the MPL above the interface inductor (5\,$\mu$m\,$\times$\,385\,$\mu$m) is reduced to 3.6\,pW by $\eta_{nf}|_{15\,\mathrm{\mu \mathrm{m}}}$ and the MPL of $10^{-3}$ transmission ($t_{_{\mathrm{MPL}},\,N=5}$ = 14\,$\mu$m). Then, the Q-factor of the resonator is maintained~\cite{PFCatJQI} (Fig.~\ref{fig_HQS} (a-ii) and (b-ii)).

The thickness of the MPL is mostly determined by a BBR wavelength. The Bragg condition of the BBR leads to a thicker layer compared to that of two-color trapping beams because of a few $\mu$m BBR peak-wavelength; the peak-wavelengths of the BBR corresponds to 9.66\,$\mathrm{\mu m}$ for 300\,K and 5.8\,$\mathrm{\mu m}$ for 500\,K. In case of no BBR consideration, the increased number of pairs for a given thickness suppresses the optical transmission more. The scattering line source can be reduced to a few\,pW level close to the level which does not affect a Q-factor~\cite{PFCatJQI} by $\eta_{wg}|_{5\,\mu\mathrm{m}}$ and the MPL of $10^{-6} \textendash 10^{-5}$ transmission ($t_{_{\mathrm{MPL},\,N=15}} = 15 \cdot t_{_{\mathrm{MPL},\,N=1}} \simeq 4.5\,\mu\mathrm{m}$) with the validation of a hybrid quantum system at $5\,\mu\mathrm{m}$ (see Fig.~\ref{fig_HQS} (a-i) and (b-i)).

The absorption-induced heat flux $\mathcal{H}_{abs}$ in a nanowaveguide is estimated from the extinction coefficient $k$ of $\mathrm{SiN_x}$. For an example, an input heat flux is $\mathcal{H}_{abs} = {P}_{0} / A_{wg} \cdot (1-\exp(-4 \pi \cdot k \cdot l_{wg} / \lambda)) = 1.57 \times 10^5 \,\textendash\,10^6\,\mathrm{W/m^2}$ with $k = 10^{-11}\,\textendash\,10^{-10}$, where the input optical power $P_{0}$ = 30\,mW, the effective mode area $A_{wg}$ = 800\,nm $\times$ 300\,nm, and a nanowaveguide length $l_{wg}$ = 1\,cm. A 300\,nm-thick $\mathrm{SiN_x}$ waveguide is deposited on $\mathrm{SiO_{2}}$ layer and Si substrate, and this sample is supposed to be anchored at the 3.5\,K cooling stage. The highest temperature of the $\mathrm{SiN_x}$ waveguide is about $4.2\,\textendash\,10.3$\,K from the numerical simulation by COMSOL$^{\textregistered}$; the heat transfer simulation considers heat conduction through the materials, and buffer-gas cooling at a pressure less than $10^{-10}$\,mbar in a dilution refrigerator can be neglected due to cryo-pumping. For a 1\,cm-long nanofiber with 500\,nm diameter, 30\,mW input power corresponds to $\sim$750\,K in a room temperature vacuum chamber~\cite{Wuttke13} and corresponds to $\sim$450\,K when non-tapered, coated fiber sections of a nanofiber are attached to a U-shaped metallic holder; the nanofiber section is 1\,cm long except linearly tapered fiber sections connected to non-tapered fiber sections, and the metallic holder is anchored at the 3.5\,K cooing stage. The heat conduction of a nanofiber is a hundred times worse than that of a $\mathrm{SiN_x}$ waveguide because of its geometry; a longer fiber attached to the holder induces more heat due to its lower heat conduction.
 
\begin{figure}
\centering\includegraphics[width=0.8\textwidth]{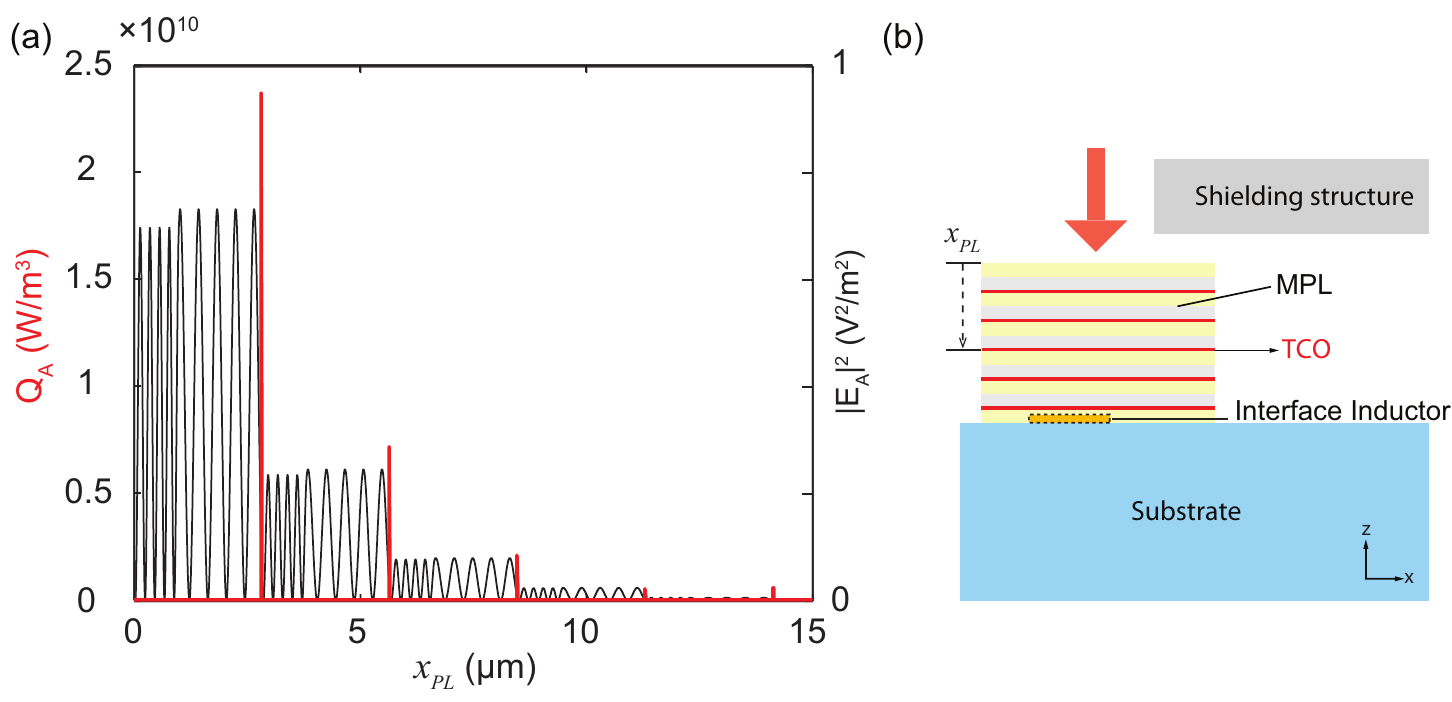}
\caption{The absorbed optical power Q per unit volume through the MPL with a thin nanogrid conductive dielectric layer ($t_{_{\mathrm{MPL},\,N = 5}} \simeq 14\,\mu\mathrm{m}$). (a) Q ($\mathrm{W/m^3}$) vs. a location $x_{PL}$ ($\mu$m) at the MPL; the absorbed power Q at TCO layers (red) and the electric field square of $|E_{A}|^2$ (black) in the MPL are represented. The absorbed power at the first TCO layer (2\,nm thin) is 240\,nW from a 1\,mW scattered light that would radiate uniformly. Note that two different oscillations in each dielectric pair are observed because of two different optical properties. (b) Cross-sectional view of design geometry (xz plane, not-to-scale).}
\label{fig_abs_Bragg}
\end{figure}

The absorbed power at the optimal MPL with a thin conductive dielectric layer is studied in Fig.~\ref{fig_abs_Bragg}. The optimal MPL can suppress the transmission of the optical fields and has broad stop bands for preventing the BBR, but the absorbed power in the MPL could generate heat in the sample stage and increase thermal background. The total heat has to be lower than the cooling power of 100\,mK cooling stage ($200\,\textendash\,400\,\mu\mathrm{W}$). From the waveguide sample, the intensity of 1\,mW line optical source is assumed to be radiated uniformly. We calculate the electric field square in the MPL and estimate the absorbed power per unit volume~\cite{YKim07} with $Q_{A} = (2\pi c \epsilon_0 n_A k_A / \lambda) |E_{A}|^2$, where $n_A$ and $k_A$ are given by $n_A + i \cdot k_A = \sqrt{\tilde{\epsilon}_{A}(\omega)}$, and $E_{A}$ is the electric field component of the optical field in the MPL. In the first TCO layer, the absorbed optical power is $P_{abs\,1st} = Q_{A} \cdot V_{A} \simeq  240\,\mathrm{nW} \ll 200\mu\mathrm{W}$, where $Q_{A}$ = $2 \times 10^{10}\,\mathrm{W/m^3}$ and the volume $V_{A}$ = $15\,\mathrm{\mu m} \times 395\,\mathrm{\mu m} \times 2\,\mathrm{nm}$ with a 2\,nm thin conductive layer. The absorbed powers of next other layers are decreased more. The total absorbed power $P_{abs,\,tot}$ through the MPL lower than the cooling power may be negligible because the equilibrium temperature will be lower than the temperature of the cooling stage. However, this absorption can create a thermal gradient in the MPL which may increase thermal background level and can affect the Q-factor by the local thermal excitation of quasiparticles. The single-photon operation requires a low thermal background level to resolve single photons.

\begin{figure}
\centering\includegraphics[width=0.8\textwidth]{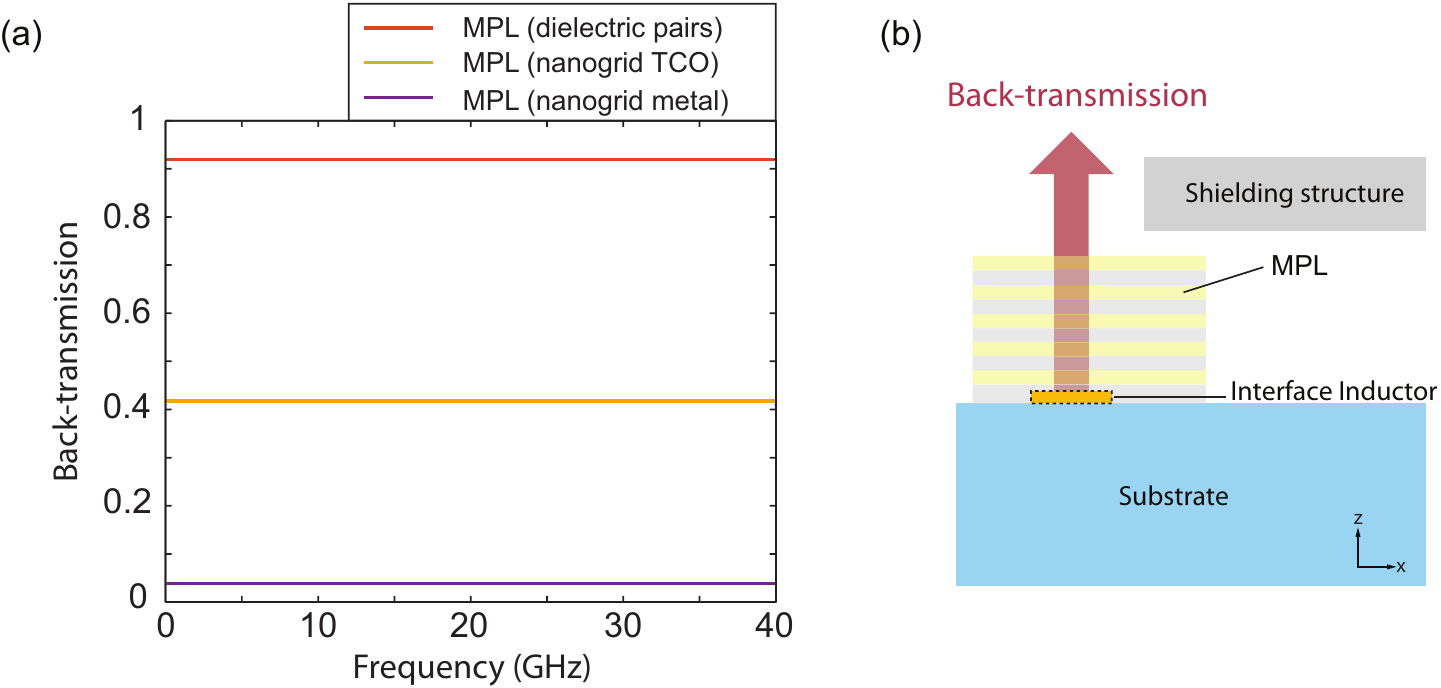}
\caption{The back-transmission of SM fields through the MPLs ($t_{_{\mathrm{MPL},\,N = 5}} \simeq 14\,\mu\mathrm{m}$). (a) The calculated back transmission shows a local field attenuation around the interface inductor with the MPL; the MPL of dielectric pairs (red), the MPL with a thin nanogrid conductive dielectric layer (yellow), and the MPL with a thin nanogrid metallic layer (purple) (b) Cross-sectional view of design geometry (xz plane, not-to-scale).}
\label{fig_back_transmission}
\end{figure}

Figure~\ref{fig_back_transmission} shows the calculated back transmissions of SM fields with the MPLs at microwave frequencies of DC to 40\,GHz, assuming the near-field AC field is a plane wave. The MPL of dielectric pairs has a low microwave attenuation less than 10\,\%, but the narrow stop-bands of the MPL of dielectric pairs cannot cover the full range of the BBR. The MPL with absorptive nanogrid metallic thin-films can suppress a broadband MIR BBR, but the attenuation of SM fields is drastically increased more than 90\% and the partially absorbed MIR and NIR energy might heat up the device. An absorptive layer with conductive dielectrics such as $\mathrm{TiO_2}$ can reduce SM fields less than 60\% while the transmissions for both NIR photons and a broadband MIR BBR can slightly increase. The calculated back transmission of SM fields represents a local field attenuation from the MPL around the interface inductor. We also made rigorous simulations of the back transmission from the near-field microwave using an FEM software to validate plane-wave approximation, which showed similar results.

\begin{figure}
\centering\includegraphics[width=0.8\textwidth]{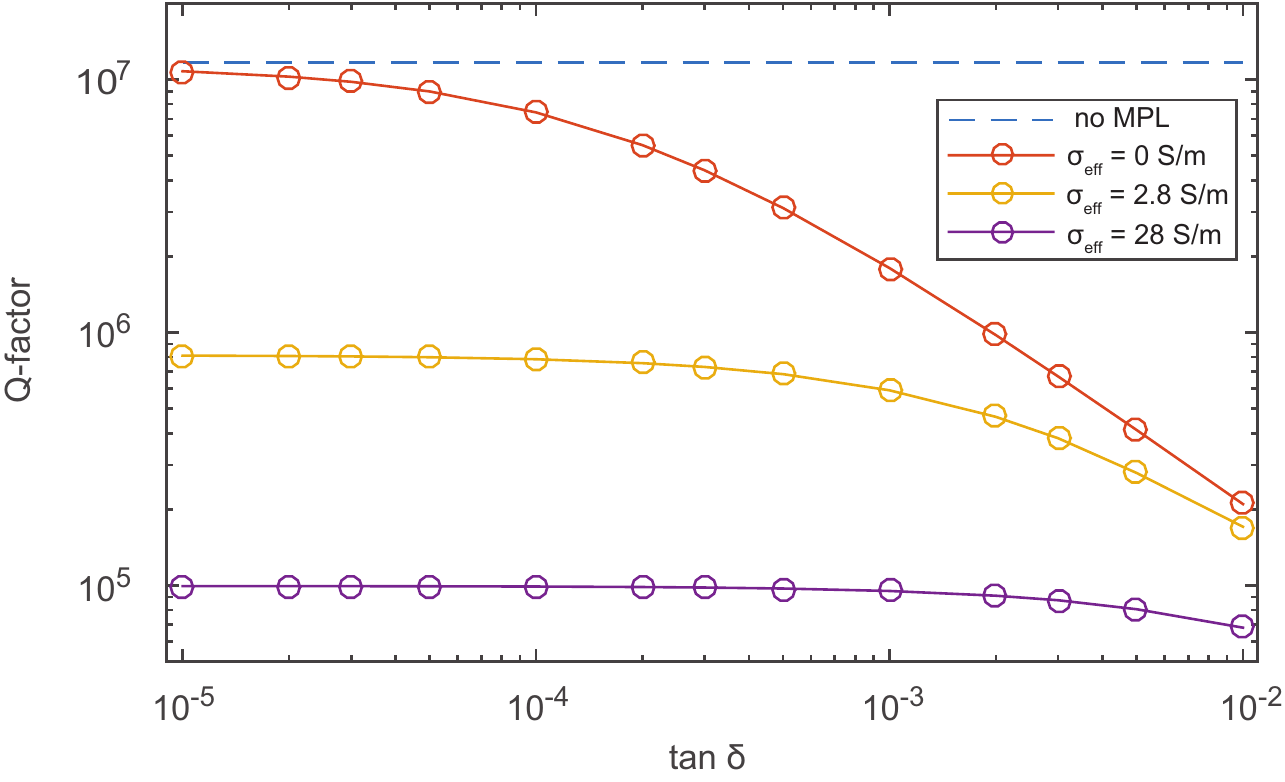}
\caption{The reduced Q-factor of a lumped-element SMR by additional TLS losses from the MPL with a thin nanogrid conductive dielectric layer ($t_{_{\mathrm{MPL},\,N = 5}} \simeq 14\,\mu\mathrm{m}$). A realistic TLS loss tangent range from the MPL is used for the single-photon regime ($T \simeq 20\,\mathrm{mK}$), where TLS losses are in excess of dielectric losses. The MPL ($15 \times 395 \times 14 \,\mu\mathrm{m}^3$) on the interface inductor minimally overlaps with the cavity mode. The effective conductivity for void structures is defined for the nanogrid conductive dielectric layer. The Q-factors of no MPL (blue dash), the MPL of dielectric pairs (red, $\sigma_{eff} = 0\,\mathrm{S/m}$), the MPL with a thin nanogrid conductive dielectric layer (yellow, $\sigma_{eff} = 2.8\,\mathrm{S/m}$), and the MPL with a thin conductive dielectric layer (purple, $\sigma_{eff} = 28\,\mathrm{S/m}$).}
\label{fig_Q_factor}
\end{figure}

Additional losses are caused by the MPLs on the interface inductor of a lumped-element SMR; the loss is minimized because a localized magnetic field is less sensitive to dielectrics and electric dipole moments and the MPL is partially overlapped with a cavity mode. In case of the single-photon regime at a low temperature ($T \simeq 20\,\mathrm{mK}$), TLS losses were measured in refs~\cite{Gordon14, Macha10, Martinis05, Pappas11b}, where TLS losses dominate dielectric losses. Based on the measurements, we chose a realistic TLS loss tangent range from the MPL. We assume that an additional TLS loss tangent from the MPL can be added to a dielectric loss tangent of dielectric pairs in the MPL and a loss tangent of a thin conductive dielectric layer can be remained as the same due to its high conductive loss. In the HFSS$\textsuperscript{\texttrademark}$ simulation (3D FEM), we regard the normalized microwave electro-magnetic field in the simulation as the SM electro-magnetic field of the single-photon regime, and we simulate reduced Q-factors semiclassically with a realistic TLS loss tangent range ($\tan \delta = 10^{-5} \sim 10^{-2}$) from the MPL and with a conductive loss from a conductive dielectric layer ($\sigma_{eff}$ = 0\,$\sim$\,28\,S/m); we assume that the conductivity of a conductive dielectric layer ($\mathrm{TiO_{2}}$) at a low temperature is comparable to one at room temperature. In case of the MPL with a thin nanogrid conductive dielectric layer, we use the first and second dielectric layers with conductive layers as follows: 0.98\,$\mu$m-thin $n_1$ dielectric layer ($\epsilon_r$ = 6, $\mu_r$ = 1, $\tan \delta$), 1.83\,$\mu$m-thin $n_2$ dielectric layer ($\epsilon_r$ = 2, $\mu_r$ = 1, $\tan \delta$), and 2\,nm-thin $n_3$ nanogrid conductive oxide layer with void structures ($\epsilon_r$ = 100, $\mu_r$ = 1, $\sigma_{eff}$ = 2.8\,S/m) are defined at the microwave regime. The void structures reducing vortex-induced loss have a lower effective conductive loss depending on the ratio of voids as given in Eq.~(\ref{Maxwell-Garnett}). The conductivity $\sigma_{eff}$ of $n_3$ layer without void structures is estimated as 28\,S/m~\cite{Chen09}. The simulation results for the reduced Q-factors of a thin-film lumped-element SMR with the MPLs is shown in Fig.~\ref{fig_Q_factor}; the Q-factor with no MPL is $1.2 \times10^7$ with the Sapphire substrate ($\tan \delta = 1 \times 10^{-7}$), and we assume that this Q-factor includes an intrinsic dielectric loss and an intrinsic TLS dissipation. We note that the MPL with $\tan \delta < 10^{-3}$ gives a still large Q-factor $>10^5$, that doesn't decrease the performance of the SMR. In addition, the increased effective conductivity of the MPL from 0 to 28\,S/m slightly reduces a resonant frequency by a couple hundred kHz, but dielectric pairs of the MPL on the interface inductor increases a resonant frequency much from 6.22\,GHz (without MPL) to 6.95\,GHz. The separate simulation shows that the increased volume of the MPL ($\sigma_{eff}$ = 2.8\,S/m, $\tan \delta = 10^{-4}$, $t_{_{\mathrm{MPL},\,N = 5}} = 14\,\mu\mathrm{m}$) from $15 \times 395 \times 14\,\mu\mathrm{m}^3$ to $50 \times 395 \times 14\,\mu\mathrm{m}^3$ reduces the Q-factor from $7.8 \times 10^5$ to $9.5 \times 10^4$ due to the increased overlapping volume between the MPL and the cavity mode.

\section{Conclusions}
We have proposed the MPL to protect a lumped-element SMR against the multiband spectrum of photons and BBR and realize a hybrid quantum system with the evanescent field atom trap. The hybrid quantum system should maintain a high Q-factor of the SMR with low dissipations in the single-photon regime. Surface-scattered NIR photons and absorption-induced broadband MIR BBR by the evanescent field atom trap reduce a Q-factor due to optically- and thermally-excited quasiparticles. The MPL was designed to suppress both quasiparticles on the interface inductor, and the floated shielding structure to protect capacitors and other inductors. We also estimated the reduced Q-factors from several types of the MPLs by finite element method with a realistic TLS loss tangent range in the single-photon regime. The MPL with a thin nanogrid conductive dielectric layer is an optimal solution for two-color NIR photons and a broadband MIR BBR, maintaining $\sim 10^5$ Q-factor with dissipations for the single-photon operation. The evanescent field atom trap such as a waveguide/nanofiber was considered for the design of the MPL. Compared to a nanofiber, a waveguide induces much lower absorption-induced BBR due to good heat conduction but higher surface-scattered photons due to its surface roughness. Compared to a waveguide, a nanofiber induces lower surface-scattered photons due to good surface roughness but higher absorption-induced BBR due to low heat conduction. Based on the validation of a hybrid quantum system, $\sim 10^5$ Q-factor necessitates $10^5 \sim 10^6$ trapped atoms, and the trapped atom number might be possible for ultracold atoms but would be challenging for cold atoms, loading into the evanescent field atom trap. Our MPL scheme can be adjusted in general depending on NIR and MIR spectrums and be a practical solution to reduce quasiparticle dissipations in a lumped-element SMR.

\section*{Acknowledgements}
This work was funded by ARO Atomtronics MURI project. We would like to thank S. L. Rolston, J. B. Hertzberg, and all PFC@JQI ``Atoms-on-SQUIDs'' team members including C. Ballard, R. P. Budoyo, K. D. Voigt, Z. Kim, J. A. Grover, J. E. Hoffman, S. Ravets, P. Solano, J. R. Anderson, C. Lobb, L. A. Orozco, F. C. Wellstood for useful discussions. We also thank J. B. Hertzberg and R. M. Lewis for their careful reading and comments.

\begin{appendix}
\section{Loss mechanism of a SMR}
\label{Appx_SC_sensitivity}

\subsection*{Quasiparticle dissipation}
The energy of NIR photons or a broad MIR BBR higher than a superconducting bandgap energy $2 \Delta$ can break Cooper pairs and create quasiparticles~\cite{Barends11}. The quasiparticle dissipation in a SMR with its density $n_{QP}$ and a resonant frequency $\nu_r$ (for $kT \ll h \nu_r$) can be expressed as
\begin{equation}
\frac{1}{Q_{QP}} = \frac{\alpha}{\pi}\sqrt{\frac{2\Delta}{h \nu_{r}}} \frac{n_{QP}}{D(E_F)\Delta},
\end{equation}
where $D(E_F)$ is the two-spin density of states, and $\alpha$ is the kinetic inductance fraction. A SMR with a bigger $\Delta$ and a higher $\nu_{r}$ becomes more protective against the quasiparticle dissipation.

Based on Mattis-Bardeen theory, quasiparticles can be diffused and recombined in a SMR. The quasiparticle dissipation can be generated from thermal excitation $G_{T}$ and optical excitation $G_{opt}$. The rate equation of the quasiparticle density $n_{QP}$ is given as
\begin{equation}
\frac{\partial n_{QP}}{\partial t} = G_{T} + G_{opt} - R \cdot n_{QP}^2 + D \cdot \nabla^2 n_{QP},
\end{equation}
where $G_{opt} \propto P_{opt}/\Delta$ with an optical power $P_{opt}$, the recombination rate $R$, and the diffusion rate $D$. For a high $P_{opt}$ at the steady state, the quasiparticle density is proportional to $\sqrt{P_{opt}/\Delta}$, where the light-induced density exceeds the thermal background.

The allowable power was measured by a fiber-coupled NIR photons on a lumped-element SMR~\cite{PFCatJQI}. Since inductors are sensitive to quasiparitcles due to kinetic inductance, the allowable power is proportional to the inductor area. The MPL improves the protection capability against quasiparticle dissipation with a floated shielding structure, and the allowable power increases. In consideration of the allowable power, thermally excited quasiparticles are similar to optically excited quasiparticles.

\subsection*{Two-level system dissipation}
The main source of TLS dissipation is hydrogen impurities forming $O\textnormal{-}H$ bonds near metal-vacuum boundary or at interfaces between a thin-film and a substrate~\cite{Gordon14}. The TLS dissipation of a superconducting coplanar waveguide resonator~\cite{Macha10} was studied at milikelvin temperature.
Below a critical power, the TLS loss rate becomes power independent, and for a fixed power the loss rate increases significantly with decreasing temperature. The temperature dependence reflects the occupation difference of the energy levels of the TLSs. The TLS loss rate~\cite{Macha10, Martinis05, Pappas11b} is expressed as:
\begin{equation}
\frac{1}{Q_{TLS}} = \frac{F \pi \rho p^2}{3\epsilon} \frac{\tanh(\frac{h \nu_{r}}{2 k_B T})}{\sqrt{1 + \left(\frac{E}{E_{c}}\right)^2}},
\end{equation}
where $\rho$ is the TLS density of states with a fluctuating dipole moment $p$; $F$ is the filling factor of the TLSs hosting medium; $(E/E_{c})^2 = (\Omega_{TLS})^2 T_{1} T_{2}$; $\Omega_{TLS} = p \cdot E / \hbar$ is the Rabi frequency proportional to the electric field $E$; $T_{1}$ and $T_{2}$ are the relaxation times of TLSs. A low power regime, $(E/E_{c})^2 \ll 1$,  is independent of the power dependence, $1/\sqrt{1 + (E/E_{c})^2} \approx 1$, which increases a loss rate, but a power beyond a low power limit ($(E/E_{c})^2 > 1$) decreases a loss rate due to the power dependence. A low temperature, $T \leq T_{c}/10$, increases a loss rate. A high temperature, but $T < T_{c}$, decreases a loss rate due to the temperature dependence of $\tanh(h \nu_{r}/(2 k_B T))$. For the realization of a hybrid quantum system, we need a low power and low temperature regime which is sensitive to TLS dissipation.

\section{The estimation of surface-scattered photons and absorption-induced blackbody radiation}
\label{Appx_Est}

\subsection*{Surface-scattered near-infrared photons}
It is well known that optical losses in waveguides consist of material (both intrinsic and extrinsic) absorption, Rayleigh scattering, and waveguide imperfections~\cite{Agrawal10}. The intrinsic absorption of a nanophotonic atom trap is a heat source because the small mode area of the nanophotonic waveguide makes localized and intense BBR source in the near field. Compared to a nanofiber atom trap~\cite{Wuttke13}, the local equilibrium temperature of the ridge waveguide can be lower due to its large substrate area and close thermal-contact with 3.5\,K cooling stage. In addition to the intrinsic absorption, the waveguide imperfections contribute direct illumination to a SMR underneath, while extrinsic absorption and Rayleigh scattering can be relatively small due to smooth surface morphology by the state-of-the-art plasma-enhanced chemical-vapour deposition (PECVD) system or low-pressure chemical vapour deposition (LPCVD) system. Intrinsic absorption is expected to be slightly higher than silica fiber ( $<$\,0.5\,dB/km~\cite{Agrawal10}) used in modern optical communication system because of the lower bandgap of $\mathrm{SiN_x}$~\cite{Acevedo13}. Assuming that the strength of the Lorentzian model for $\mathrm{SiN_x}$ is comparable with silica, the loss is estimated to be less than 5\,dB/km at the wavelength of 750\,nm and 1064\,nm~\cite{Scopel02}.  In fact, $\mathrm{SiN_x}$ waveguide with propagation losses less than 0.1\,dB/cm at the wavelength of 632\,nm had been reported in 1977~\cite{Streifer77} and Gondarenko et. al demonstrated 0.055\,dB/cm at the telecommunication wavelength~\cite{Lipson09}. A major source of the propagation loss is waveguide imperfections such as surface roughness and an refraction index difference of core and cladding layers. An evanescent field usually operates as strongly-guided modes, where their refraction index difference and propagation loss are higher than those of weakly-guided modes.

We show the surface scattering loss from the waveguide~\cite{Payne94, Barwicz05, Ciminelli09} in order to estimate the absorbed optical power per unit volume at the MPL on the SMRs as shown in Fig.~\ref{fig_abs_Bragg}. Based on Ref.~\cite{Barwicz05}, the scattering losses normalized to the roughness variance of $\mathrm{SiN_x}$ / $\mathrm{SiO_2}$ waveguides (TE-like mode) is 0.01\,(dB/cm)/$\mathrm{nm^2}$ for 800\,nm width (core) and 300\,nm height (core), where $n_{\mathrm{core}}$ = 2.00 and $n_{\mathrm{clad}}$ = 1.45 for 1550\,nm. After considering the wavelength-dependent scattering losses ($\alpha_{sc} \propto 1/\lambda^{m}$ (dB/cm) with $m \simeq 4.2$ for the TE-like mode~\cite{Ciminelli09}), the scattering losses for 1064\,nm light is 0.05\,(dB/cm)/$\mathrm{nm^2}$. Therefore, the roughness variance of the waveguide determines the scattering losses. If the top-surface of the waveguide has 1\,$\mathrm{nm^2}$ roughness variance, the scattering losses from the top-surface is 0.05\,(dB/cm).

Dominant scattering results from the side-wall roughness because of the imperfection of photolithographic waveguide patterns on photoresist before the etching process. In Fig.~\ref{fig_concept}, direct illumination from the top surface of deposited $\mathrm{SiN_x}$ affects SMRs underneath the waveguide structure. From planar-waveguide loss theory~\cite{Payne94} (2D analysis), illuminated power can be estimated based on the surface roughness and the correlation length of the roughness. The scattering losses of the planar waveguide can be modeled more reliably in 3D model, and the loss analysis in 3D is an order of magnitude smaller than the loss analysis in 2D~\cite{Barwicz05, Ciminelli09}.

\subsection*{Absorption-induced heat and mid-infrared blackbody radiation}
For a total optical trapping power of $P_0$,  the absorbed optical power in the $\mathrm{SiN_x}$ can be $P_0 \cdot (1 - \exp(-4 \pi \cdot k \cdot l_{wg} / \lambda ))$ with the length of a waveguide $l_{wg}$ and the extinction coefficient $k = 10^{-11}$ despite a low intrinsic absorption loss $k \sim 10^{-12}$ from Ref.~\cite{Agrawal10}. The extinction coefficient $k$ partially resulting in joule heating, excluding Rayleigh scattering portion, is assumed to be less than $10^{-11}$ at the wavelengths of our interest, based on the total optical losses. We simulate the thermal gradient with this input heat flux in COMSOL$^{\textregistered}$ 2D simulation with boundary condition (the waveguide sample anchored at 3.5\,K cooling stage). This local heat source creates the BBR which should be prevented to protect SMRs from thermally-excited quasiparticles. The BBR spectrum is defined as $S_{BBR}(\nu,T) = (2 h \nu^3 / c^2) \cdot 1/(\exp\left(h\nu/(k_B T)\right)-1)$, where a peak-spectrum frequency of $S_{BBR}(\nu_{pk},T$) is $\nu_{BBR}(T) \approx 2.82 \cdot k_B T / h$.

\end{appendix}


\begin{thebibliography}{99}
\bibitem{PFCatJQI} A lumped-element aluminum SMR without MPLs can maintain a Q-factor ($\simeq 5 \times 10^5$) from a fiber-coupled NIR photons if the intensity is less than $1 \times 10^{-4}\,\mathrm{pW/\mu m^2}$.
\end{thebibliography}
\end{document}